# A Case Study on Virtual and Physical I/O Throughputs

*By Dr. Timur Mirzoev, Dr. Baijian Yang, Mr. Marcus Davis, & Mr. Travis Williams*

# A Case Study on Virtual and Physical I/O Throughputs

*By Dr. Timur Mirzoev, Dr. Baijian Yang, Mr. Marcus Davis, & Mr. Travis Williams*

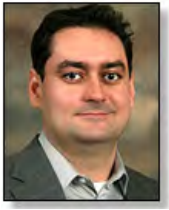
Dr. Timur Mirzoev is a professor of Information Technology Department at Georgia Southern University, College of Information Technology. Dr. Mirzoev heads the International VMware IT Academy Center and EMC Academic Alliance at Georgia Southern University. Some of Dr. Mirzoev's research interests include server and network storage virtualization, information systems, storage networks and topologies. Currently, Dr. Mirzoev is holds the following certifications: VMware Certified Instructor, VMware Certified Professional 4, EMC Proven Professional, LefthandNetworks (HP), SAN/iQ, A+. Timur Mirzoev can be contacted at tmirzoev@georgiasouthern.edu.

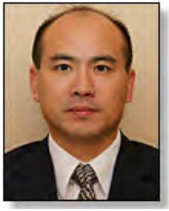
Dr. Baijian Yang is currently an Associate Professor at the Department of Technology, Ball State University. He has extensive industry and academic experiences in mobile computing, distributed computing, and information security. His current industry certifications include MCSE, CISSP, and Six Sigma Black Belt. Dr. Yang is also the contributing author of books 'Professional Smartphone Programming', and 'Windows Phone 7 Programming for Android and iOS Developers' by Wiley/WROX. Dr. Yang received his Ph.D. in Computer Science from Michigan State University in 2002. Baijian Yang can be contacted at byang@bsu.edu.

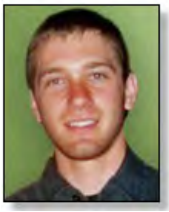
Mr. Marcus Davis is a Master of Science in Information Technology scholar at Carnegie Mellon University's Australian campus. Mr. Davis serves as the President and Founder of Kinetic Influence LLC, an IT consulting firm based in Atlanta, Ga. His research interests include global distributed systems, cloud based infrastructure virtualization and human computer interaction. Mr. Davis is the recipient of Carnegie Mellon's Scholarship for Emerging Leaders and has been recognized by the University of Rochester's Xerox Award for Innovation in Information Technology. Marcus Davis can be contacted at mdavis27@georgiasouthern.edu.

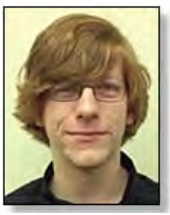
Mr. Travis Williams is currently a part time student at Georgia Southern University and full time employee as a Senior Solutions Developer for Morris Technology. Travis is just finishing up his 3rd year at Georgia Southern University as an Information Technology major. Travis has been on the President's and Dean's list and was recently inducted into the Honor's Society Upsilon Pi Epsilon. After graduating from Georgia Southern, Travis hopes to find a job as a Network Administrator and put the knowledge he has gained to use. Travis Williams can be contacted at tw01049@georgiasouthern.edu.

## ABSTRACT
Input/Output (I/O) performance is one of the key areas that need to be carefully examined to better support IT services. With the rapid development and deployment of virtualization technology, many essential business applications have been migrated to the virtualized platform due to reduced cost and improved agility. However, the impact of such transition on the I/O performance is not very well studied. In this research project, the authors investigated the disk write request performance on a virtual storage interface and on a physical storage interface. Specifically, the study aimed to identify whether a virtual SCSI disk controller can process 4KB and 32KB I/O write requests faster than a standard physical IDE controller. The experiments of this study were constructed in a way to best emulate real world IT configurations. The results were carefully analyzed. The results reveal that a virtual SCSI controller can process smaller write requests (4KB) faster than the physical IDE controller but it is outperformed by its physical counterpart if the sizes of write request are bigger (32KB). This manuscript presents the details of this research along with recommendations for improving virtual I/O performance.

## INTRODUCTION
Five decades of hard drive technology is now the industry standard for data storage. One variant of the technology known as Small Computer System Interface (SCSI) has become commonplace in enterprise settings. The separation of commands used to control the SCSI hard disk and the interconnect used to carry these commands creates a generic interface for the hard drive. Many interconnects have emerged to span this gap between the physical disk and the controller (Goldner, 2003).

One of the most well-known Storage Area Network (SAN) SCSI interfaces is the Fiber Channel Protocol which uses fiber optics to connect a SCSI disk array with host computers. A relatively new interconnect known as iSCSI uses an IP network to control a SCSI array (Thompson, 2002). Since iSCSI can use existing network infrastructure, it is scalable and requires less hardware when compared to other SANs (Cormier, 2008). The iSCSI technology transfer SCSI requests using TCP frames (Shrivastava & Somasundaram, 2009). "The logical link that carries the commands and data to and from TCP/IP end-points is called an iSCSI session (Hufferd, 2003). Some systems such as Openfiler are open source allowing for easy adoption (Childers, 2009). These factors make iSCSI a financially attractive storage solution for virtualization applications.

This manuscript examines the performance of physical and virtual disks controllers to compare Input/Output Operations per second (IOps) for write requests. Research on virtual disk controllers has been limited. This suggests that the novelty factor of this new technology may contribute to the absence of research on IOps for virtual systems. Several options exist in configurations of virtual storage controllers. However, the purpose of this study was to identify whether virtual SCSI disk controller processes I/O write requests of 4KB and 32KB in sizes faster than a standard physical Integrated Drive Electronics (IDE) controller. The analyzed 4KB and 32KB writes were the actual block sizes of the tested file system. Typically, if a file system is configured to use a smaller block size, it will increase the disk utilization but will negatively impact the disk I/O performance when handling large files.

## REVIEW OF RELATED LITERATURE

In recent years as hardware virtualization and high availability systems gained popularity iSCSI has become an accepted storage solution. Hypervisors such as VMware's ESX server natively support iSCSI (VMware, 2007). These systems require scalable, fast storage in order to support a virtual infrastructure. "I/O is more important than ever now that multiple virtualized operating system instances are relying on the same disk array" (Diskeeper, 2006). I/O throughput on standalone SCSI disks can be easily measured using utilities such as Iometer which was used in this study to measure write IOPs and write MB/s (Iometer, 2003).

The diverse uses of iSCSI and virtualized disks present some difficulties when trying to compare the I/O performance of physical and virtual write speeds. Write speed may be affected by the number of virtual machines running on a SAN. Differences in disk and bus speed may also affect these comparisons (Asaro, 2009). To better understand the differences between physical storage controllers and virtual storage controllers, an experiment was constructed to compare a physical Windows Server 2003 machine with a virtualized Windows Server 2003 machine located on a test Openfiler iSCSI array. The array was located in an isolated network environment to ensure that network traffic did not interfere with write requests.

With physical systems, "disk I/O is the primary performance bottleneck for a wide range of workloads" (Bhadkamkar *et al.,* 2007). When systems are virtualized, certain configurations and settings need to be in place in order to avoid bottlenecks at storage levels. Self-optimized storage systems, such as BORG – Block-reORGanization (Bhadkamkar *et al.,* 2007), attempt to improve the disk I/O performance by dynamically adjusting the block size of a file system.

A file system's block size is an essential element that allows storage system op-timization (Ponniah, 2001). Each block size contains a header which must be read before the actual data is read. If a piece of information is stored in too many blocks, the file system's performance dramatically decreases, wasting time to respond to users' requests. This is particularly true for database related applications. If a customer's record can fit into one block, then only one header is read and all of the customer's records are fetched (Ponniah, 2001). To save read/write processing time, block size may be increased to accommodate more records into one block of data on a storage device. Incorrectly set block size for storage systems hosting databases may lead to a significant loss of time during multiple read/write operations. Many Database Management Systems (DBMS) allow manual setting for block sizes that allow a reservation of a certain space inside a block for database records to expand in the future (Ponniah, 2001). Figure 1 depicts the important differences between various block size allocations for the same file.

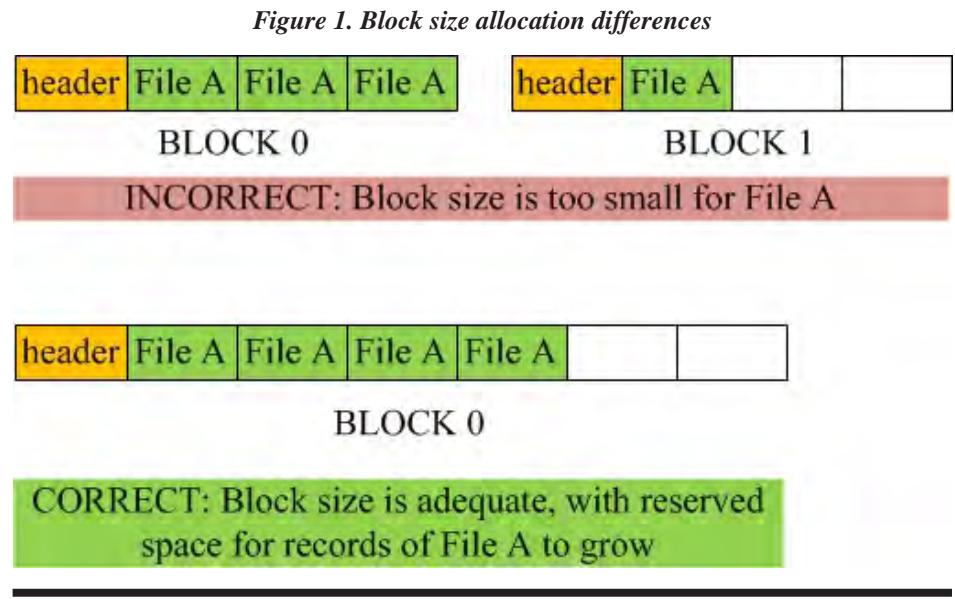

*Figure 1. Block size allocation differences*

Block size setting is closely related to a maximum and a minimum file sizes. For example, if a user requests to write a 4KB file and the storage's block size is only 2KB, then two blocks will be occupied for the same file and two headers for each block must be read for a single file. Administrators are encouraged to set the block size of a file system based on the maximum size of files to be written to storage. In Oracle databases, the most commonly used block size is 8KB which is "more than sufficient for many very large databases" (Caffrey *et al.*, 2010). In case when database records are updated and they do not fit into the original block on storage system, the entire updated record is moved to another block (Ponniah, 2001). With thousands of records and sometimes even millions of records, storage system will take a very long time to update, move, read or write records if the block size is not fit for the application.

According to Foot (2004), the performance of a database may improve once the block size on a file system is increased. With records of 8KB in size, 16KB or 32KB block sizes are recommended.

Today, many database systems have been moved to virtual environments where a database will reside on a virtual server instead of a physical server. If an administrator decides to convert a DBMS computer to a virtual machine (VM), storage settings must be meticulously set. In virtualized environments, such as our laboratory setup, disk I/O requests from a virtual machine are processed by a Linux kernel. It is the responsibility of the VMKERNEL (in our setup) to process all reads and writes on block storage servers. VMKERNEL may also create a so-called Raw Device Mapping (RDM) where, VMKERNEL creates a pointer to a raw Logical Unit Number (LUN, a storage system partition). In RDM's case, a *virtual machine processes all reads and writes on storage, not the VMKERNEL* (VMware iSCSI, 2010). Figure 2 shows an example of RDM.

For the experiments in this study, RDM was not used and all the I/O requests from virtual machines were processed by VMKERNEL.

## RESEARCH OBJECTIVES AND QUESTIONS

The purpose of this study was to identify whether virtual SCSI disk controller processes I/O write requests of 4KB and 32KB in sizes faster than a standard physical IDE controller. An additional goal was to examine the possible storage configurations for virtual machines. The substantive research questions were as follows:

$Q_1$: Does a virtual SCSI controller process 4KB I/O write requests faster than an IDE controller?

$Q_2$: Does a virtual SCSI controller process 32KB I/O write requests faster than an IDE controller?

$Q_3$: Does a virtual SCSI controller write more data than an IDE controller while processing 4KB I/O?

$Q_4$: Does a virtual SCSI controller write more data than an IDE controller while processing 32KB I/O?

**Hypotheses**

In order to find answers to the research questions, the following hypotheses were established for this study:

1. $H1_0$: $\mu_{4ph} = \mu_{4vm}$

There are no differences in IOps for write requests of 4KB between physical and virtual disk while other hardware components and software remain constant.
$H1_a$: $\mu_{4ph} \neq \mu_{4vm}$

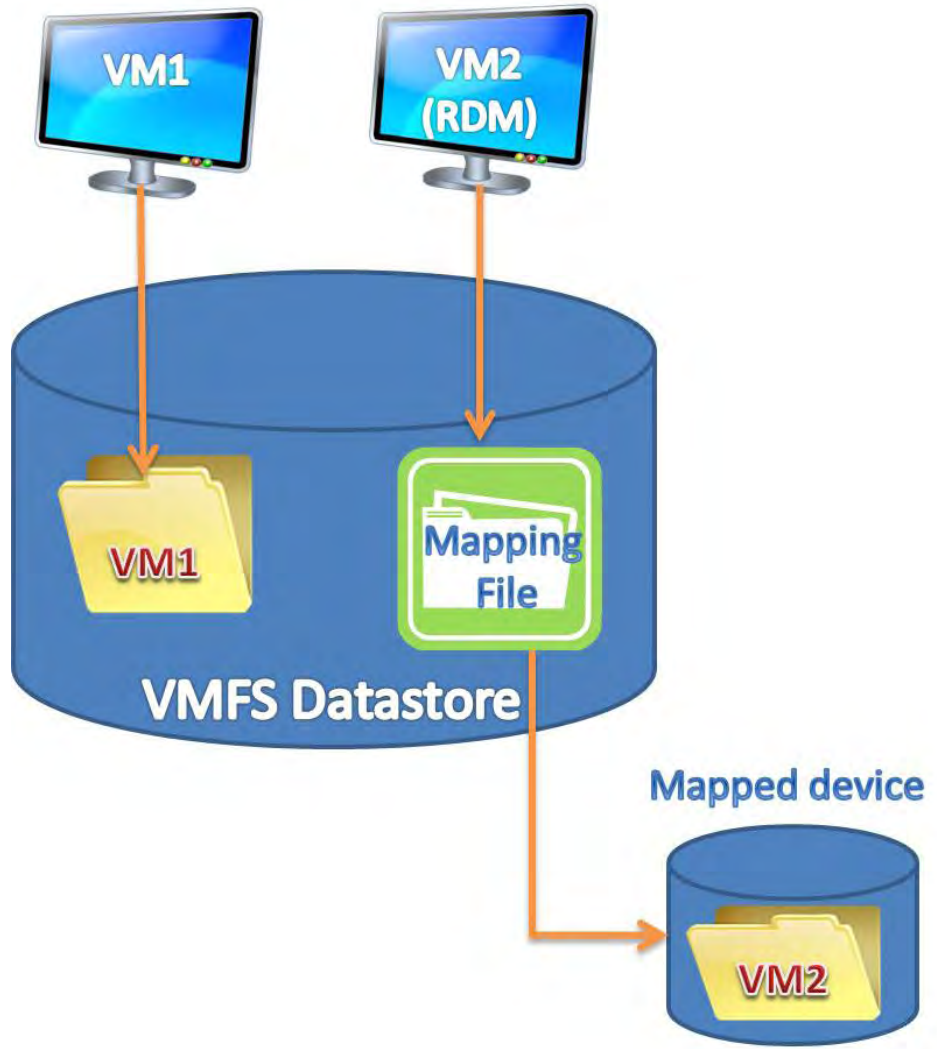

*Figure 2. An example of Raw Device Mapping*
Adapted from: VMware vSphere 4.1: Install, Configure, Manage course – Revision A

There are differences in IOps for write requests of 4KB between physical and virtual disk while other hardware components and software remain constant.

2. $H2_0$: $\mu_{32ph} = \mu_{32vm}$

There are no differences in IOps for write requests of 32KB between physical and virtual disk while other hardware components and software remain constant.

$H2_a$: $\mu_{32ph} \neq \mu_{32vm}$

There are differences in IOps for write requests of 32KB between physical and virtual disk while other hardware components and software remain constant. This study was not concerned with processor speed or memory size (although RAM is set at 2GB for each computer), the number of hosts actively using the network, the network layout or any other general aspect of the network, and the read write speed of the hard disk drives. This study was conducted to determine while in a Windows Server 2003 environment whether a virtual SCSI disk controller processes I/O write requests of 4KB and 32KB faster than a standard physical IDE controller. During this experiment, iSCSI target presented by Openfiler was exclusively connected to the ESX server with no other servers having access to the same iSCSI target on which virtual machine files resided. Fibre Channel, Content-Addressable Storage (CAS), Direct-Attached Storage (DAS) and other types of storage networks were not tested under this study.

## METHODS OF INVESTIGATION

### Research Design
In order to compare I/O write request differences between virtual and physical disks, two scenarios were created and they are presented in Table 1.

Each Iometer test ran for 40-50 minutes (5 seconds * 500 workers = 41.6 minutes) and 500 samples were collected for each size 4KB and 32KB, totaling 2000 write requests for the entire study.

### Techniques of Data Gathering
The following resources where used during this study:
1. ESX Server (version 3.5 update 2) based on IBM x336 server with a Windows Server 2003 virtual machine (VM),
2. iSCSI target presented by Openfiler NAS/SAN appliance (version 2.3),
3. Dell Optiplex GX260 with the following specifications:
   - 2 GHz Intel Processor
   - 2 Gigabytes DDR Memory
   - 80 GB Hitachi DeskStar 7200RPMs with 133MB/s IDE controller,
4. Iometer as the single IOps measuring software were used and then the results were compared,
5. Microsoft Excel and Minitab 15.0 were used to collect data and perform statistical calculations.

Iometer was used to perform all the necessary write requests on both systems. The setups for the tests were the same on each system, physical and virtual:

- Disk Target: C Drive
- Network Target: Not used
- Access Specifications: One set of 500 workers running 4K, 100% Write, 100% Sequential, and 100% Access Specification and a second set of 500 workers running 32K, 100% Write, 100% Sequential, and 100% Access specifications. Refers to *Figure* 3.
- Cycle Workers- Step Workers one at a time to all targets.
- Each test ran for 5 seconds
- Once each system's Iometer was setup to run under the appropriate specifications the tests were started and each test ran for 40-50 minutes (5 seconds * 500 workers = 41.6 minutes). Two tests were conducted for each machine: with 4KB and 32KB packet sizes.

### Openfiler setup
Openfiler NAS/SAN appliance was used to provide iSCSI target to the ESX server. ESX server connected to Logical Unit Number (LUN) 0 on OF2 (openfiler2), and that is where the virtual machine files resided, including the virtual disk of the tested virtual machine.

*Table 1. Configuration scenarios for write requests*

| Configuration | Block size, KB | |
|---|---|---|
| 1 Physical IDE | 4KB | 32KB |
| 2 Virtual SCSI | 4KB | 32KB |

It is important to understand the relevance of this discussion about Openfiler system since it processes all write requests for the virtual machines' disks. For the iSCSI target setup default values for iSCSI target under Openfiler NAS/SAN appliance were used, as shown in *Table 2*.

VMware states that the following switches need to be specified for all iSCSI targets accessed by ESX – no_root_squash, sw, sync (VMware, 2007). In this study the specifications of the recommended switches for Openfiler were not required since the NAS/SAN appliance presented targets correctly.

I/O throughput for the virtual disk of tested VM is highly dependent on the type of network that is used. Since iSCSI technology allows transmission of SCSI commands over TCP, network speed is an essential factor for iSCSI communications (Shrivastava & Somasundaram, 2009). In this study, 1000 MB/s Netgear Ethernet switch was used to connect the ESX server and the Openfiler appliance. The ESX server software iSCSI initiator was used to interconnect with the Openfiler iSCSI target. *Figure 4* presents the laboratory setup.

*Figure 3. Iometer Setup*

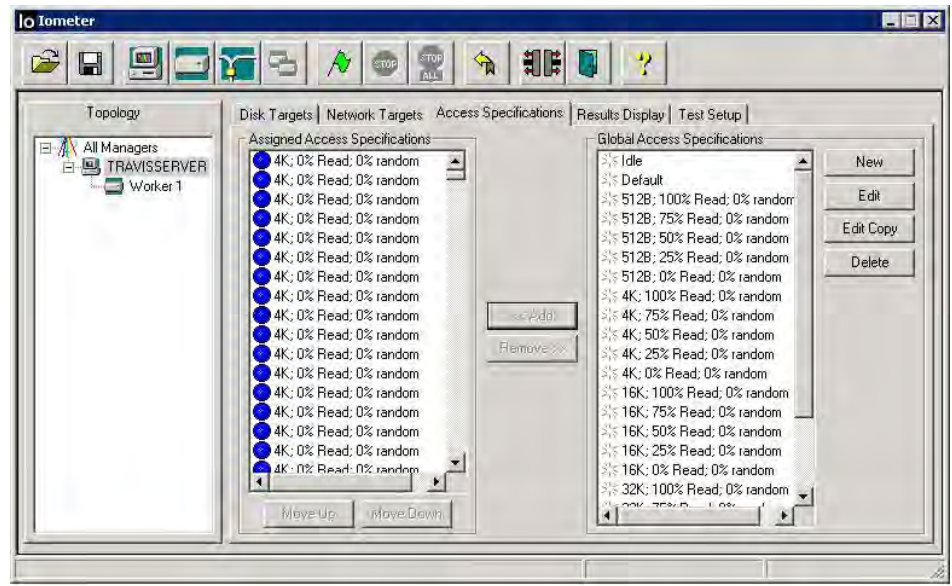

*Table 2 Openfiler default settings for iSCSI target*

| HeaderDigest | None | MaxBurstLength | 262144 | MaxRecvDataSegmentLength | 131072 |
|---|---|---|---|---|---|
| DataDigest | None | FirstBurstLength | 262144 | MaxXmitDataSegmentLength | 131072 |
| InitialR2T | Yes | DefaultTime2Wait | 2 | MaxOutstandingR2T | 8 |
| ImmediateData | No | DefaultTime2Retain | 20 | DataPDUInOrder | Yes |
| MaxConnections | 1 | Wthreads | 16 | DataSequenceInOrder | Yes |
|  |  | QueuedCommands | 32 | ErrorRecoveryLevel | 0 |

**Data Analysis Method**
The following tools were utilized for the analyses of the research questions:
1. In this study a 2-Sample t-test was used to perform a hypothesis test and compute a confidence interval of the difference between two population means of the physical and virtual PCs.
2. Minitab version 15.0 software package was used for statistical calculations.
3. Microsoft Excel software package was used for the creation of the population sample.

## FINDINGS
**Findings and Analysis of Write I/O requests**
During the simulation testing, 500 groups of data were collected. Basic descriptive statistics are then computed using Minitab software version 15.0 and results are shown in Table 3. Numeric results describe on average how many IOPs were performed on both Physical and Virtual interface for 4KB block, and 32KB block, respectively.

Histograms with normal curve for each tested configuration are shown in Figure 5 and 6.

From statistics point of view, the data collected can be characterized as an independent two-sample *t* test. Let

*Table 3. Descriptive statistics for 1 and 2 configurations*

|  | Size (KB) | Mean (IOPs) | SE Mean | St. Dev | Variance | Median (IOPs) | Range (IOPs) |
|---|---|---|---|---|---|---|---|
| 1 Physical | 4 | 875.65 | 0.82 | 18.34 | 336.25 | 873.8 | 125.18 |
|  | 32 | 881.03 | 0.871 | 19.47 | 379 | 883.08 | 122.58 |
| 2 Virtual | 4 | 1059.7 | 1.44 | 32.1 | 1032.9 | 1061.1 | 362.6 |
|  | 32 | 297.13 | 0.289 | 6.46 | 41.68 | 297.62 | 87.31 |

*Figure 4. Laboratory setup.*

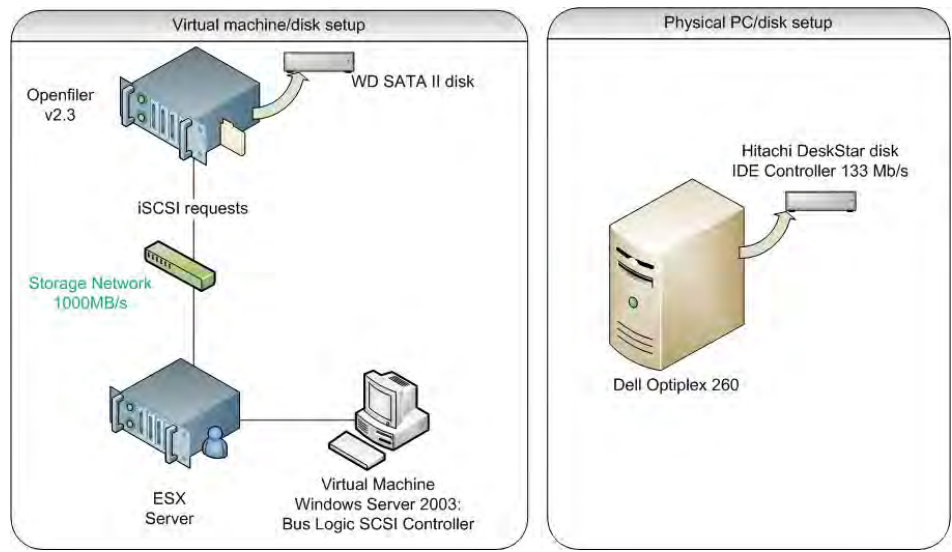

*Figure 5. Histogram for Physical Machine size*

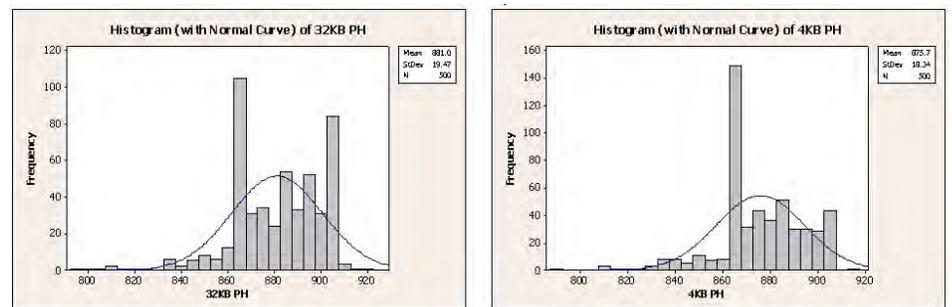

5.a. 32KB        5.b. 4KB

$\bar{x}_{4k1}$ = the mean of sample write IOPs of 4K block on the physical IDE interface

$\bar{x}_{4k2}$ = the mean of sample write IOPs of 4K block on the virtual iSCSI interface

$s_{4k1}$ = the standard deviation of sample IOPs of 4K block on the physical IDE interface

$s_{4k2}$ = the standard deviation of sample IOPs of 4K block on the virtual iSCSI interface

$n_{4k1}$ = the sample size of IOPs of 4K block on the physical IDE interface

$n_{4k2}$ = the sample size of IOPs of 4K block on the virtual iSCSI interface

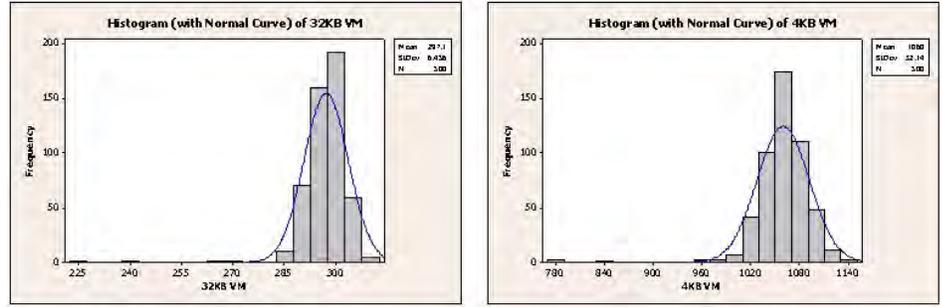

*Figure 6. Histogram for Virtual Machine*

6.a. 32KB        6.b. 4KB

From *Table* 3, it is clear that:

$\bar{x}_{4k1}$ = 875.65    $s_{4k1}$ = 18.34    $n_{4k1}$ = 500

$\bar{x}_{4k2}$ = 1059.7    $s_{4k2}$ = 32.1    $n_{4k2}$ = 500

Pooled variance is preferred over the non-pooled variance because it has more degrees of freedom and will produce more convincing results. But pooled variance should be applied only when the standard deviations of the two groups are about the same. The following formula is used to determine if the standard deviations of the two populations are equal by comparing the ratio:

$$\frac{\max(s_{4k1}, s_{4k2})}{\min(s_{4k1}, s_{4k2})} = \frac{32.1}{18.34} = 1.75 < 2$$

Since the ratio is less than 2, it is reasonable to assume that the standard deviations are equal and the pooled confidence interval procedure can be used to further analyze the data. The value of the estimated pooled standard deviation is

$$s_p = \sqrt{\frac{(n_{4k1}-1)s_{4k1}^2 + (n_{4k2}-1)s_{4k2}^2}{n_{4k1}+n_{4k2}-2}} = \sqrt{\frac{499(18.34)^2 + 499(32.1)^2}{998}} = \sqrt{\frac{682016}{998}} = 26.14$$

Thus,

$$s_p\sqrt{\frac{1}{n_{4k1}} + \frac{1}{n_{4k2}}} = 26.14\sqrt{\frac{1}{500} + \frac{1}{500}} = 1.653$$

The number of degrees of freedom for pooled standard deviation is:

$$df = n_{4k1} + n_{4k2} - 2 = 500 + 500 - 2 = 998$$

If the confidence level is set to 99%, then area in each tail is 0.01/2 = 0.005. The *t* value corresponding to a tail area of 0.005 and df = 998 is 2.581. Therefore, at 99% confidence level, the difference of two IOPs is:

$$(\bar{x}_{4k1} - \bar{x}_{4k2}) \pm ts_p\sqrt{\frac{1}{n_{4k1}} + \frac{1}{n_{4k2}}} = (875.65 - 1059.7) \pm 2.581*1.653 = -184.05 \pm 4.266$$
$$= (-188.316, -179.784)$$

Because both the lower limit and upper limit of the intervals are negative, it suggests that for the 4KB write request, at 99% confidence level the mean IOPS of the physical IDE controller is about 179.784 to 188.316 less than that of the virtual iSCSI controller.

Similarly for the 32K block size, let

$\bar{x}_{32k1}$ = the mean of sample write IOPs of 32K block on the physical IDE interface

$\bar{x}_{32k2}$ = the mean of sample write IOPs of 32K block on the virtual iSCSI interface

$s_{32k1}$ = the standard deviation of sample IOPs of 32K block on the physical IDE interface

$s_{32k2}$ = the standard deviation of sample IOPs of 32K block on the virtual iSCSI interface

$n_{32k1}$ = the sample size of IOPs of 32K block on the physical IDE interface

$n_{32k2}$ = the sample size of IOPs of 32K block on the virtual iSCSI interface

From table 2, we know that:

$\bar{x}_{32k1}$ = 881.03    $s_{32k1}$ = 19.47    $n_{32k1}$ = 500

$\bar{x}_{32k2}$ = 297.13    $s_{32k2}$ = 6.46    $n_{32k1}$ = 500

The ratio of the standard deviations is as follows:

$$\frac{\max(s_{32k1}, s_{32k2})}{\min(s_{32k1}, s_{32k2})} = \frac{19.47}{6.46} = 3.014 > 2$$

Because the ratio is greater than two, it is not reasonable to assume that the standard deviations are equal. Therefore it is necessary to follow the procedure of non-pooled confidence interval.

The point of estimate is given by

$$(\bar{x}_{32k1} - \bar{x}_{32k2}) = 881.03 - 297.13 = 583.9$$

The degrees of freedom can be approximated by
$df = min(n_{32k1} - 1, n_{32k2} - 1) = 499$.

If confidence level is still set at 99%, then the area in each tail is 0.005. From *t* distribution, when df=499, the *t* value corresponding to a tail area of 0.005 is 2.586.
Then the difference of the two 32KB write IOPS is:

$$(\bar{x}_{32k1} - \bar{x}_{32k2}) \pm t \sqrt{\frac{s_{32k1}^2}{n_{32k1}} + \frac{s_{32k2}^2}{n_{32k2}}} = 583.9 \pm 2.586 \sqrt{\frac{19.47^2}{500} + \frac{6.46^2}{500}} = 583.9 \pm 2.586\sqrt{0.842}$$
$$= 583.9 \pm 2.372 = (581.528, 586.27)$$

Since both upper limit and lower limit are positive, it means that for the 32KB write request, at 99% confidence level, the IOPs of physical IDE controller is between 581.528 and 586.27 more than that of virtual iSCSI controller.
When considering the results for Mega Byte Process Speed (MBPS), a 2 x 2 analysis of variance (ANOVA) model were used to evaluate if MBPS differed across two levels of disk controller type and data block sizes. Test of simple main effects was also performed to locate specific group differences. The level of significance was adjusted using the Bonferroni correction procedure and was accepted *a priori* at $P < 0.025$.

A significant disk controller by data block interaction was also observed ($F_{1,998} = 18787.04$, $P < 0.0001$). Simple main effects testing revealed that MBPS was significantly larger for the Physical Disk controller (27.4±0.6) compared to the Virtual Disk controller (4.1±0.12) for the 4KB data block size ($P < 0.0001$). For the 32 KB block size, MBPS was also significantly larger for the Physical Disk controller (27.5±0.61) compared to the Virtual Disk controller (9.29±0.20; $P < 0.0001$). When assessing MBPS across Data Block Size for the Virtual Disk controller, MBPS was significantly larger for the 32 KB (9.29±0.20) compared to the 4 KB size (4.14±0.13; $P < 0.0001$; similarly, for the Physical Disk controller MBPS values were also significantly larger for the 32 KB size (27.5±0.61) compared to the 4 KB size (27.4±0.57; $P < 0.0001$).

Based on the test results and analyses, the following is concluded:

1. $H1_0$: $\mu_{4ph} = \mu_{4vm}$ is rejected in favor of the alternative $H1_a$: $\mu_{4ph} \neq \mu_{4vm}$: there are differences in IOps for write requests of 4KB between physical and virtual disk while other hardware components and software remain constant.

2. $H2_0$: $\mu_{32ph} = \mu_{32vm}$ is rejected in favor of the alternative $H2_a$: $\mu_{32ph} \neq \mu_{32vm}$: there are differences in IOps for write requests of 32KB between physical and virtual disk while other hardware components and software remain constant.
This study's research questions were answered in the following manner:

$Q_1$: Does virtual SCSI controller processes 4KB I/O write requests faster than IDE controller?
 *Yes, virtual SCSI controller processes 4KB I/O write requests faster than IDE controller.*
$Q_2$: Does virtual SCSI controller processes 32KB I/O write requests faster than IDE controller?
 *No, virtual SCSI controller does not process 32KB I/O write requests faster than IDE controller.*
$Q_3$: Does virtual SCSI controller write more data than IDE controller while processing 4KB I/O?
 *No, virtual SCSI controller does not write more data than IDE controller while processing 4KB I/O.*
$Q_4$: Does virtual SCSI controller write more data than IDE controller while processing 32KB I/O?
 *No, virtual SCSI controller does not write more data than IDE controller while processing 32KB I/O.*

## LIMITATIONS OF THE STUDY AND FUTURE CONSIDERATIONS

Due to budget and time restrictions, the data collected in this study is not comprehensive. The testing environment was set up to resemble typical storage systems in real life scenarios. However, the results of write performance may vary significantly if different hardware systems or software systems are applied. So changes in the instrument (hardware), benchmarking software, write block sizes, and the duration of the tests may jeopardize internal validity of this study.

In particular, following important factors that were not examined under this research could alter the performance of a virtual SCSI controller. Further investigations are needed in future works.

1. Type of virtual disk being created – independent persistent or non-persistent disk (VMware, 2007).
2. Type of network interface cards (NIC) for the hypervisor (ESX server in this study) that provide the connectivity to iSCSI targets. For example it is possible that a 10 GB/s NIC will process write/read requests faster than 1000 MB/s NIC that was used in this study.
3. Use of hardware SCSI host bus adapter (HBA) for the hypervisor may allow for processing write/read requests

faster than a built-in software iSCSI initiator as was used in this study.
4. Provisioning of iSCSI targets may be approached differently from the approach used for this study. In addition to Openfiler, other iSCSI storage can be used to better understand the performance differences.

In addition, this study by itself is not sufficient to explain the results we have observed. The authors believe it may due to the fact that the block size of the physical SCSI hard drive is also small. Therefore, the I/O performance of a virtual SCSI interface is good when the size of the request is 4KB (small). And the I/O performance declines significantly when the size of the request is 32KB (large): additional overheard occurs to map a virtual block to a few physical blocks. Another possible cause of the phenomena could be the size of the disk write buffer. If the disk write buffer is big enough to hold the sequential 4KB write requests but not enough to contain the sequential 32KB write requests, then it can explain why the virtual SCSI interface can achieve great I/O performance when the requests are small and poor performance when the requests are large. More experiments are therefore needed to discover the root cause of problem.

## *CONCLUSIONS*
Based on the findings of this study, it is apparent that in a typical storage system, smaller size write requests are processed quicker via virtual SCSI controller. This result suggests that it is more efficient to deploy a VM with a virtual SCSI controller for frequent, small write requests. An example of such demand could be a database that constantly creates small size write requests. For instance, a database that is configured to store records of security auditing log needs to be updated very often and each record is typically very small in size. As for larger write requests, such as 32KB, a physical hard disk controller is more capable in comparison with a virtual storage controller.

There are some changes that could be made to improve the performance of a virtual storage controller. According to Perilli(2005), the following is recommended for better I/O performance when deploying a virtual machine: a) create a dedicated partition for virtual machines only, b) create guest virtual disks with "Allocate all disk space now" option, and c) schedule a daily defragmentation for the virtual machines' directories. It is important to understand, that while the virtual SCSI controller processes write requests faster, it does not write more data in either case of 4KB or 32KB requests. This phenomenon maybe explained by the fact that the underlying hardware for the virtual controller is a network storage device, that initially has to be reached by the ESX host and only after the connection is established, can the data be written to the disk. A reasonable solution for a case where the virtual controller processes I/O faster than the physical controller but the amount of data written by a virtual controller is not larger than the physical controller's, is to use a write-back type of cache on storage devices. In such devices a write acknowledgement is sent as soon as cache memory receives a write request and the actual recording of data is processed at a later time (Shrivastava & Somasundaram, 2009). Database systems may benefit from such setup.

In conclusion, it is important to consider performing more tests with virtual systems as they overwhelmingly surmount enterprise environments. With introductions of new storage technologies based on 40 GB/s or even 100 GB/s systems (Dornan, 2008), there may be a significant improvement in performance of virtual storage controllers.

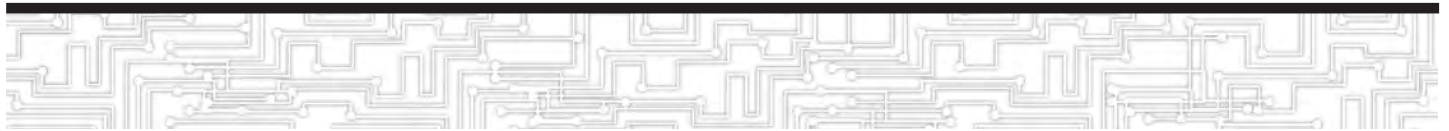